\documentclass[aps,prb,twocolumn,a4paper,10pt]{revtex4-1} 

\usepackage[T1]{fontenc}		
\usepackage[english]{babel}		
\usepackage[latin1]{inputenc}	
\usepackage{times}

\usepackage{amsmath}			
\usepackage{amssymb}			
\usepackage{bbm}				

\usepackage[colorlinks=true, a4paper=true, pdfstartview=FitV,linkcolor=blue, citecolor=blue, urlcolor=blue]{hyperref}

\usepackage{graphicx}			
\usepackage{graphics}			
\DeclareGraphicsExtensions{.pdf}
\usepackage{color}		

\newcommand{\bea}{\begin{eqnarray}}
\newcommand{\eea}{\end{eqnarray}}

\newcommand{\mbf}{\mathbf}
\newcommand{\mc}{\mathcal}

\newcommand{\df}{\check}

\newcommand{\Eqref}[1]{Eq.~\eqref{#1}}
\newcommand{\Figref}[1]{Fig.~\ref{#1}}
\newcommand{\ie}{{\it i.e.~}} 

\newcommand{\gsphasediff}{\bar\varphi_{ij}}
\newcommand{\Vector}[1]{\left( \begin{array}{c}#1\end{array}\right)}
\newcommand{\MatrixTwo}[1]{\left( \begin{array}{c c}#1\end{array}\right)}
\newcommand{\MatrixThree}[1]{\left( \begin{array}{c c c}#1\end{array}\right)}

\begin{document}
\title{Length scales, collective modes, and type-1.5 regimes in three-band superconductors }

\author{Johan~Carlstr\"om${}^{1,2}$,~Julien~Garaud${}^{2}$,~and~Egor~Babaev${}^{1,2}$}
\affiliation{ 
${}^1$ Department of Theoretical Physics, The Royal Institute of Technology, Stockholm, SE-10691 Sweden \\
${}^2$ Department of Physics, University of Massachusetts Amherst, MA 01003 USA 
}

\date{\today}

\begin{abstract}
The recent discovery of iron pnictide superconductors has resulted in a rapidly growing interest 
in multiband models with more than two bands. In this work we specifically focus on the properties of 
three-band Ginzburg-Landau models which do not have direct counterparts in  more 
studied two-band models. First we derive normal modes and characteristic length scales in the
conventional $U(1)$ three-band Ginzburg-Landau model as well as in its time reversal symmetry 
broken counterpart with $U(1)\times Z_2$ symmetry. We show that in the latter case, the normal modes 
are mixed phase/density collective excitations. A possibility of the appearance of a
massless phase-difference mode associated with fluctuations of the phase difference is also discussed.
Next we show that gradients of densities and phase differences can be inextricably intertwined 
in vortex excitations in three-band models. 
This can lead to very long-range attractive intervortex interactions and 
appearance of type-1.5
regimes even when the intercomponent Josephson coupling is large. In some cases it also results in the 
formation of a domain-like structures in the form of a ring of suppressed density around a vortex across which one of the phases shifts by $\pi$.
We also show that field-induced vortices can lead to a change of broken symmetry from $U(1)$ to 
$U(1)\times Z_2$ in the system. In the type-1.5 regime, it results in a semi-Meissner state where the system 
has a macroscopic phase separation in domains with broken $U(1)$ and $U(1)\times Z_2$ symmetries.
\end{abstract}

\maketitle


Superconductivity with two gaps associated with different bands was  first theoretically predicted in 1959 \cite{suhl,*moskalenko}.
However  it was not until 42 years later, with the discovery of $MgB_2$ \cite{mgb2} that it started to attract a wide interest 
(for a recent review see \cite{xi}).
Because the condensates in two band superconductors are not independently
conserved, the  system considered in \cite{suhl,*moskalenko} share 
the same broken $U(1)$ symmetry of the ground state as their single-component counterparts.
The interband tunneling results in a system that attains its free energy minimum when the phase difference between 
the condensates is either zero or $\pi$.
Nonetheless in 1969  it was discussed that individual phases of the two condensate wave functions are 
important degrees of freedom, since they give rise  to a new kind of collective excitations. These collective 
excitations are associated with the 
fluctuations of the relative phase of the two superconducting components  around its ground state value: 
 the Leggett's mode \cite{leggett}, for a recent discussion see \cite{sharapov}. A report of the observation of the 
Leggett's  mode in $MgB_2$ appeared very recently \cite{blumberg}.

Another example of new physics which can arise in two-band system (as compared to their single-band counterparts) is 
associated with disparity of the characteristic length scales of density variations. That is, a single quantum vortex 
in a two-band system, should in general produce two different cores. 
As a consequence of this, there appears a regime which was recently termed type-1.5 superconductivity \cite{1type-15}.
In that  regime the two characteristic length scales of density variations $\xi_1$ and $\xi_2$ satisfy the 
condition $\xi_1<\sqrt{2} \lambda <\xi_2$. 
For a subset of parameters in this regime there are thermodynamically 
stable vortices with non-monotonic interaction. Namely,
 these vortices exhibit interaction which is long range attractive, and short-range repulsive. 
As a consequence of the long range attraction between vortices the system allows an additional ``semi-Meissner" phase associated 
with a macroscopic phase separations in domains of the Meissner state and vortex states see e.g.
\cite{bs1,1type-15,GL2mass,2type-15,geurts,nonpairwise,Silaev,daonew}.
For a detailed introduction see  \cite{GL2mass2}.

In the last three years there has been a rapidly growing interest  in  multiband superconductivity with more than two 
components. The interest was sparked by the recent discovery of iron pnictide superconductors \cite{iron1}
and  subsequent discussions that superconductivity in these systems may be described by a theory  with more than two 
relevant bands \cite{iron2,iron3}. 
It was observed that  the inclusion of a third band in the theory in several respects leads 
to qualitatively different physics compared to two-band systems \cite{nagaosa,stanev,tanaka,hu}. The new physics arises from the fact 
that the presence of three or more components can lead to phase frustration. It results from competition of 
three or more interband Josephson coupling terms, which cannot all simultaneously attain the most energetically 
favorable phase locking pattern \cite{nagaosa,stanev,tanaka,hu}. This frustration leads to Time Reversal Symmetry 
Breakdown (TRSB)
\cite{agterberg,nagaosa,stanev,tanaka,hu} (we discuss it more quantitatively below).
See also \cite{zhang,zhang2} for a different discussion of possible time reversal symmetry breakdown in iron pnictides. 
Here we show that phase frustration leads to a plethora of new phenomena in 
the physics of collective excitations and the magnetic response 
of the three-band Ginzburg-Landau model.

In the TRSB 
phase there are no ``phase-only" Leggett's modes. Instead there is a different kind of collective excitations: Mixed phase-density modes.
These mixed normal 
modes have quite complex structure and, they can possess modes with large characteristic length scales even in the case of strong 
Josephson coupling. 
At the transition point to the TRSB regime,
the length scale of one of the phase difference modes diverges, rendering
one of the modes massless (as was also discussed recently
in a London model \cite{hu2}, for other recent discussions of 
Leggett's modes in connection with iron pnictides see \cite{machida,meera}). Note however that if the phase transition in the TRSB state is first order as was argued in Ref. \cite{stanev}, then there will be no massless mode.
 This is in contrast to two-band systems where increasing interband Josephson coupling always 
diminishes disparities of the density variations \cite{GL2mass,GL2mass2}. In particular it implies that the 
type-1.5 regime is possible in three band superconductors even in cases of quite strong 
interband Josephson coupling. Moreover we show that 
in three-band systems the Semi-Meissner state can represent not 
only a macroscopic phase separation in vortex and Meissner domains but also represent a macroscopic phase separations 
of domains with different broken symmetries.

\section{Model}
The minimal GL free energy functional to model a three-band superconductor  is
\begin{align} 
 F= &\frac{1}{2}(\nabla \times \mbf A)^2+ \sum_{i =1,2,3}\frac{1}{2}|\mbf D\psi_i|^2 
+\alpha_i|\psi_i|^2+\frac{1}{2}\beta_i|\psi_i|^4 \nonumber\\
+&\sum_{i =1,2,3}\sum_{j>i}\eta_{ij}|\psi_i||\psi_j|\cos(\varphi_{ij}) \,.
\label{freeEnergy}
\end{align}
Here $\mbf D=\nabla+ie\mbf A$ and $\psi_i=|\psi_i| e^{i\varphi}$ 
are complex fields representing the superconducting components. 
The phase difference between two condensates is denoted $\varphi_{ij}=\varphi_j-\varphi_i$.
The magnetic flux density through the system is given by $\mbf B=(\nabla\times \mbf A)$ and the magnetic energy 
density is $\mbf B^2/2$.
Such a multicomponent GL free energy can in certain cases be microscopically derived
at temperatures close but not too close to $T_c$ (for a review see \cite{gurevich}).
Indeed the existence of three superconducting bands is not by any means a
sufficient condition for a system to have GL expansion like that given in \Eqref{freeEnergy}.
 However many of the questions which we consider below in fact  do not require 
the system to be in the high-temperature region where a GL expansion like \Eqref{freeEnergy} could
in certain cases be
formally  justified. In what follows we will however use  the minimal GL model 
since it provides a convenient framework to discuss this physics qualitatively.
  In the \Eqref{freeEnergy} the coefficients $\alpha_i$ change signs at some characteristic temperatures which are generally 
different for all components. Below this temperature $\alpha_i<0$ and the band is active. Above it, $\alpha_i>0$ and the band is 
passive. Passive bands can nevertheless have nonzero superfluid density because of the interband Josephson tunneling 
terms $\eta_{ij}|\psi_i||\psi_j|\cos\varphi_{ij}$. Thus it is possible in this model to 
have only passive bands, and still nonzero superfluid densities due to Josephson terms. In the three component 
model \Eqref{freeEnergy} there are additional terms allowed by symmetry, e.g. bi-quadratic terms in density 
(for a review of microscopic derivation of such terms from a weak-coupling two-band theory see \cite{gurevich}). 
However the impact of these terms on length scales and vortex physics in three-band model is essentially the same as in 
the well studied two-band case \cite{GL2mass2}. Since their role is mostly connected with a straightforward renormalization 
of the length scales we will not repeat this analysis here. Instead we will focus primarily on the  Josephson couplings, which can 
play principally different roles in two- and three- band cases.

Let us first discuss the simplest London approximation \ie $|\psi|=const$. Then one can extract gradients of the 
gauge-invariant phase differences by rewriting the model as:
\begin{align} 
 F= &\frac{1}{2\sum_{i =1,2,3}|\psi_i|^2}{[\sum_{i =1,2,3}|\psi_i|^2\nabla\varphi_i +e\sum_{i =1,2,3}|\psi_i|^2 {\bf A}]^2}\nonumber \\
+&\frac{|\psi_1|^2|\psi_2|^2}{2\sum_{i =1,2,3}|\psi_i|^2}(\nabla (\varphi_1-\varphi_2))^2
 \nonumber\\
 +&\frac{|\psi_2|^2|\psi_3|^2}{2\sum_{i =1,2,3}|\psi_i|^2}(\nabla (\varphi_2-\varphi_3))^2
 \nonumber\\
 +&\frac{|\psi_1|^2|\psi_3|^2}{2\sum_{i =1,2,3}|\psi_i|^2}(\nabla (\varphi_1-\varphi_3))^2
 \nonumber\\
+&\sum_{i =1,2,3}\sum_{j>i}\eta_{ij}|\psi_i||\psi_j|\cos(\varphi_i-\varphi_j) \nonumber\\
+&\frac{1}{2}( \nabla \times \mbf A)^2  
\label{LondonEnergy}
\end{align}
The first term features the phase gradients coupled to the vector potential: this corresponds to the total current in the system.
The rest of the terms correspond to counter-flow of carriers in different bands. Since there is no charge transfer in 
counter-flows there is no coupling to gauge fields. In the limit $\eta_{ij}=0$ the second, third and fourth term describe neutral
superfluid modes with phase stiffnesses ${|\psi_i|^2|\psi_j|^2}/[{2\sum_{i =1,2,3}|\psi_i|^2}]$ which were
studied in detail in \cite{smiseth}. When Josephson terms are present they break symmetry by giving preferred 
values to the phase differences, yet the system can have fluctuations near these values. After this illustration
of phase fluctuations, we discuss in the following the fluctuations within the full Ginzburg-Landau
model which involves fluctuations of both phases and densities.

 Systems with more than two Josephson-coupled bands can exhibit 
 \emph{phase frustration}. For $\eta_{ij}<0$, a given Josephson interaction energy term is minimal for zero phase difference
 (we then refer to the coupling as ``phase-locking" ), while when  $\eta_{ij}>0$ it is minimal for a phase difference equal to $\pi$ 
 (we then refer to the coupling as ``phase-antilocking" ). Two component systems are symmetric with respect to the sign change 
$\eta_{ij}\to -\eta_{ij}$ as the phase difference changes by a factor $\pi$, for the system to recover the same interaction. However, in 
systems with more than two bands there is generally no such symmetry. For example if a three band system has $\eta>0$ for all 
Josephson interactions, then these terms can not be simultaneously minimized, as this would correspond to all 
possible phase differences being equal to $\pi$. 

\section{Ground state of a three band superconductor}
The ground state values of the fields  $|\psi_i|$ and $\varphi_{ij}$ of system \Eqref{freeEnergy} are found by minimizing its potential 
energy 
\bea
\sum_i\Big\{\alpha_i|\psi_i|^2+\frac{1}{2}\beta_i|\psi_i|^4\Big\}
+\sum_{j>i}\eta_{ij}|\psi_i||\psi_j|\cos(\varphi_{ij}).
\label{potential}
\eea
Minimizing the potential energy \Eqref{potential} can not in general be done analytically. Yet, some properties 
can be derived from qualitative arguments. In terms of the sign of the $\eta$'s, there are four principal situations: 

\begin{center}
\begin{tabular}{ c||c|cc } 
Case & Sign of $\eta_{12},\eta_{13},\eta_{23}$ & Ground State Phases \\ 
\hline
1& $- - -$ & $\varphi_1=\varphi_2=\varphi_3$ \\
2& $- - +$ & Frustrated  \\ 
3& $- + +$ & $\varphi_1=\varphi_2=\varphi_3+\pi$ \\
4& $+ + +$ & Frustrated
 \end{tabular}
\end{center}

The case 2) can result in several ground states. If $|\eta_{23}|\ll |\eta_{12}|,\;|\eta_{13}|$, then the phase differences 
are generally $\varphi_{ij}=0$. If on the other hand $|\eta_{12}|,\;|\eta_{13}| \ll |\eta_{23}| $ then $\varphi_{23}=\pi$ 
and $\varphi_{12}$ is either $0$ or $\pi$. For certain parameter values it can also have compromise states with 
$\varphi_{ij}$ not being integer multiples of $\pi$. 

\begin{figure}[!htb]
 \hbox to \linewidth{ \hss
\includegraphics[width=\linewidth]{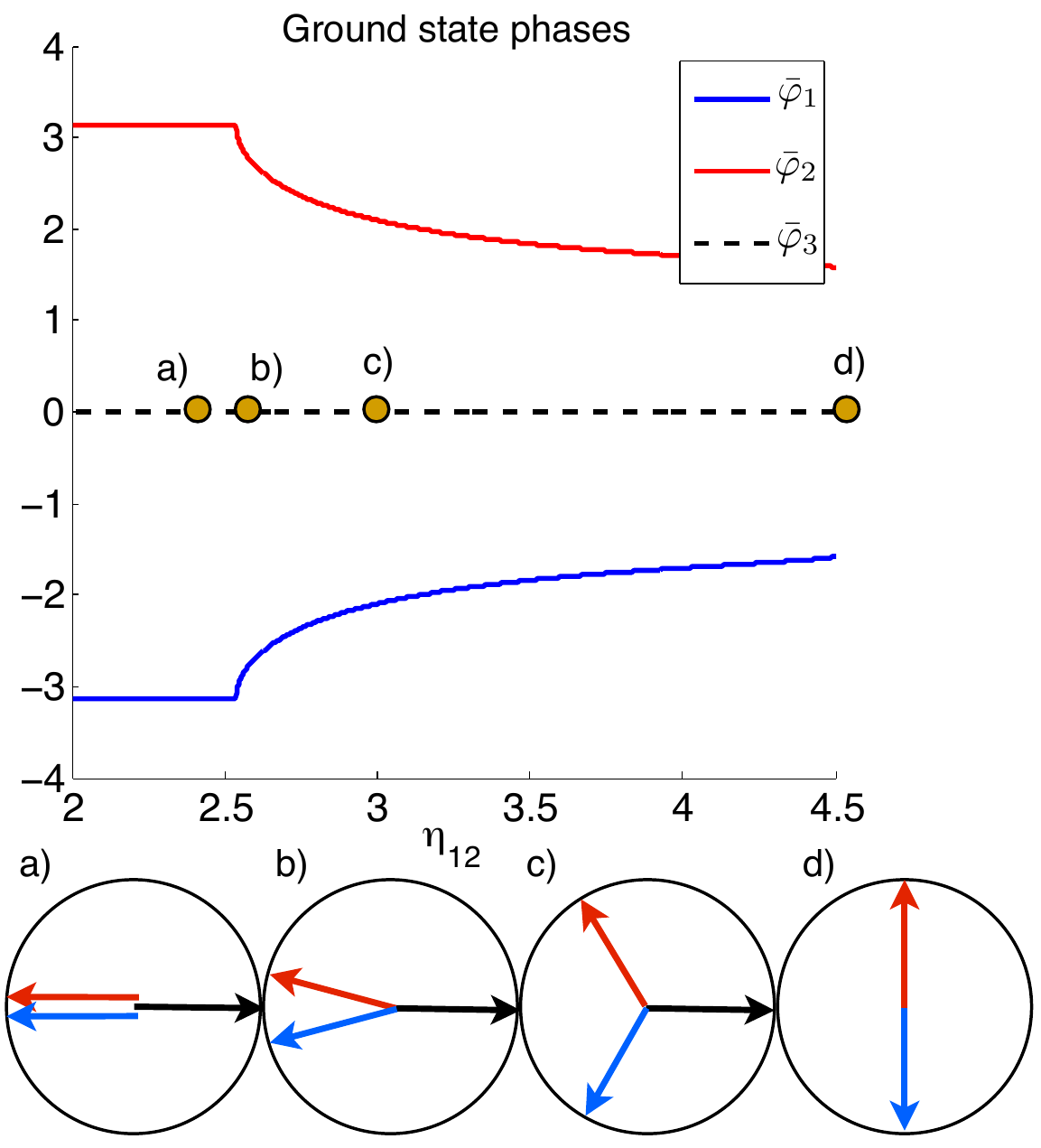}
 \hss}
\caption{
Ground state phases of the three components as function of $\eta_{12}$ (here $\varphi_3=0$ fixes the gauge). 
The GL parameters are $\alpha_i=1,\;\beta_i=1,\;\eta_{13}=\eta_{23}=3$. For intermediate values of $\eta_{12}$ the ground state 
exhibits discrete degeneracy (symmetry is $U(1)\times Z_2$ rather than $U(1)$) since the energy is invariant under the sign 
change $\varphi_2\to-\varphi_2,\; \varphi_3\to-\varphi_3$. For large $\eta_{12}$ we get $\varphi_2-\varphi_3=\pi$ 
implying that $|\psi_3|=0$ and so there is a second transition from $U(1)\times Z_2$ to $U(1)$ and only two bands at the point d). 
Here, the phases were computed in a system with only passive bands, though systems with active bands exhibit the same 
qualitative properties except for the transition to $U(1)$ and two bands only (\ie active bands have non-zero density in the ground state).  
}
\label{case4}
\end{figure}

The case 4) can give a wide range of ground states, as can be seen in \Figref{case4}. As $\eta_{12}$ is scaled, 
ground state phases change continuously from  $( -\pi,\; \pi,\; 0)$ to the limit where one band is depleted and 
the remaining phases are $(-\pi/2,\;\pi/2)$. 

An important property of the potential energy \Eqref{potential} is that it is invariant under complex conjugation
of the fields. That is, the potential energy does not change if the sign of all phase differences is changed,  
$\varphi_{ij}\to-\varphi_{ij}$. Thus, if any of the phase differences $\varphi_{ij}$ is not an integer multiple of $\pi$, 
then the ground state posses an additional discrete $Z_2$ degeneracy. For example for a system with 
$\alpha_i=-1,\;\beta_i=1$ and $\eta_{ij}=1$, two possible ground state are given by 
$\varphi_{12}=2\pi/3,\;\varphi_{13}=-2\pi/3$ or $\varphi_{12}=-2\pi/3,\;\varphi_{13}=2\pi/3$.
 Thus in this case, the symmetry  is $U(1)\times Z_2$, as opposed to $U(1)$.
As a result, like any other system with $Z_2$  degeneracy, the theory allows an additional set of topological excitations : 
domain walls interpolating between the two inequivalent ground states. Under certain conditions the system also 
does allow composite topological excitations which are bound states of closed domain walls and vortices \cite{skyrmions}.

We are interested in determining quantitatively (i) the ground state densities and phase differences and (ii) the characteristic 
length scales at which a perturbed field recovers its ground state values. These quantities are derived from a perturbative 
expansion around the ground state. Consider the following expansion of the fields entering the Ginzburg-Landau 
free energy functional \Eqref{freeEnergy}, around the ground state
\begin{align}
  \psi_i&=(u_i+\epsilon_i(r))\exp\left\lbrace i(\bar\varphi_i+\phi_i(r)) \right\rbrace\,, \nonumber \\
   \mbf{A}&=\left(\frac{a(r)}{r}\right)(\sin\theta,\cos\theta)\,.
\label{Expansion}
\end{align}
The ground state densities and phases are denoted  $u_i$ and $\bar\varphi_i$ respectively. Since we are interested in vortex 
excitations, lets  consider an axially symmetric configuration by requiring that the field fluctuations $\epsilon_i(r)$, $\phi_i(r)$ 
and $a(r)$, depend only on the radial coordinate. The expansion \Eqref{Expansion} is inserted into the free energy  \Eqref{freeEnergy} which is then sorted 
by growing orders in the fluctuations, namely $F=F^{(0)}+F^{(1)}+F^{(2)}+\dots$. The condensation energy is given by $F^{(0)}$. 

\subsection{Ground state}

The ground state can be represented by the vector of the zeroth order degrees of freedom of \Eqref{Expansion}, 
\bea
\gamma^{(0)}=( u_1,u_2,u_3,\bar\varphi_1,\bar\varphi_2,\bar\varphi_3)^T.
\eea
The fluctuation amplitudes are collected in the $6$ entry vector
\bea
\gamma^{(1)}=( \epsilon_1,\epsilon_2,\epsilon_3,\phi_1,\phi_2,\phi_3)^T.
\eea
The gauge field fluctuation $a$ decouples from the other fluctuations. The term in the GL free energy 
which is linear in the fluctuations reads
\begin{align}
 F^{(1)}&=\sum_{i}2u_i\epsilon_i(\alpha_i+\beta_iu_i^2)
+\sum_{j>i}\eta_{ij}(u_i\epsilon_j+u_j\epsilon_i)\cos\gsphasediff			\nonumber \\
+&\sum_{j>i}\eta_{ij} u_iu_j(\phi_j-\phi_i)\sin\gsphasediff \,.\,
\label{order1}
\end{align}
where $\gsphasediff$ denote phase differences of the ground state. \Eqref{order1} is a linear (in the fluctuations) 
system of $6$ equations. Since we consider fluctuations near the ground state it has to be zero for any arbitrary 
fluctuation. Indeed, by definition, no fluctuation can decrease the  energy of the ground state. Positive definiteness 
implies that all the prefactors of the fluctuations are zero. Thus expanding  \Eqref{order1} and collecting the 
prefactors of the fluctuation amplitudes gives the system of $6$ equations which determine the ground state vector 
$\gamma^{(0)}=( u_1,u_2,u_3,\bar\varphi_1,\bar\varphi_2,\bar\varphi_3)^T$. The system reads explicitly
\begin{subequations}\label{GSequations}
\begin{align}
0&= \alpha_1u_1+\beta_1u_1^3+\frac{\eta_{12}}{2}u_2\cos\bar\varphi_{12} +\frac{\eta_{13}}{2}u_3\cos\bar\varphi_{13}	\label{GSequation1} \\
0&= \alpha_2u_2+\beta_2u_2^3+\frac{\eta_{12}}{2}u_1\cos\bar\varphi_{12} +\frac{\eta_{23}}{2}u_3\cos\bar\varphi_{23}	\label{GSequation2} \\
0&= \alpha_3u_3+\beta_3u_3^3+\frac{\eta_{13}}{2}u_1\cos\bar\varphi_{13} +\frac{\eta_{23}}{2}u_2\cos\bar\varphi_{23} 	\label{GSequation3} \\
0&= -\eta_{12}u_1u_2\sin\bar\varphi_{12} -\eta_{13}u_1u_3\sin\bar\varphi_{13}	\label{GSequation4} \\
0&= \eta_{12}u_1u_2\sin\bar\varphi_{12} -\eta_{23}u_2u_3\sin\bar\varphi_{23}	\label{GSequation5} \\
0&= \eta_{13}u_1u_3\sin\bar\varphi_{13} +\eta_{23}u_2u_3\sin\bar\varphi_{23}	\label{GSequation6} \,.
\end{align}
\end{subequations}
Except under very specific conditions this cannot be solved analytically. In this paper we aim at the most general 
structure of the ground state, so no further assumptions will be made and the problem is solved using 
numerical methods (we here used Newton-Raphson algorithm).
For numerical calculations of the ground state values of the fields, it is convenient to fix the gauge by for 
example imposing $\bar\varphi_1=0$. 

\subsection{Length scales}

Once the ground state $\gamma^{(0)}$ is known, relevant information about the physics of the system can be extracted from the 
quadratic order $F^{(2)}$ of the fluctuation expansion (note that this is equivalent to considering linearized GL equations).
The fluctuations are described by a system of Klein-Gordon equations for  the $6$ condensate fluctuations 
($3$ densities plus $3$ phases), supplemented by a Proca field equation which describes fluctuations of 
the gauge field. For studying the system it may be convenient to switch to a slightly different basis 
\bea
\gamma^{(1)}=( \epsilon_1,\epsilon_2,\epsilon_3,\pi_1,\pi_2,\pi_3)^T~\text{where}~\phi_i\equiv\frac{\pi_i}{u_i}
\eea
since in this basis, the (squared) mass matrix of the Klein-Gordon system is symmetric. 
The results will be straightforwardly switched back to the basis $\phi$.
The total functional at this order reads 
\bea
F^{(2)}=E_{\mbox{\tiny Klein-Gordon}}+E_{\mbox{\tiny Proca}}\,,
\eea
where   
\begin{align}
E_{\mbox{\tiny Klein-Gordon}} &\equiv  \frac{1}{2}(\gamma^{(1)\,\prime})^2 +\gamma^{(1)}\mathcal{M}^2\gamma^{(1)}\nonumber \\
E_{\mbox{\tiny Proca}} &\equiv  \frac{1}{2}(\frac{a^\prime}{r})^2+\frac{e^2}{2r^2}\sum_iu_i^2  a^2 \,.
\label{Proca}
\end{align}
Here ${}^\prime$ denotes the differentiation with respect to the radial coordinate $r$. 
The (squared) mass matrix $\mathcal{M}^2$ of the Klein-Gordon system can easily be read from 
\begin{align}
 \gamma^{(1)}\mathcal{M}^2\gamma^{(1)}=&\sum_{i}\epsilon_i^2(\alpha_i+3\beta_iu_i^2)+\sum_{j>i}\eta_{ij}\epsilon_i\epsilon_j\cos\gsphasediff\nonumber \\
+&\sum_{j>i}\eta_{ij}\Big\lbrace(u_i\epsilon_j+u_j\epsilon_i)\left(\frac{\pi_j}{u_j}-\frac{\pi_i}{u_i}\right)\sin\gsphasediff\nonumber\\
&\ \ \ \ \ \ \ \ \ \ \ \ \  -\frac{u_iu_j}{2}\left(\frac{\pi_j}{u_j}-\frac{\pi_i}{u_i}\right)^2\cos\gsphasediff\Big\rbrace \,,
\label{order2}
\end{align}
simply by identifying the prefactors of the perturbations and filling the corresponding entries in the mass matrix. 
Before discussing in detail this mass matrix, let us consider the Proca equation, for the mass of the gauge field. 
It is the easiest length scale to derive, since the Proca equation for the gauge field fluctuation \Eqref{Proca} decouples 
from all other. The London penetration depth of the magnetic field $\lambda$ is the inverse mass of the gauge field, namely 
\begin{equation}
 \lambda\equiv\frac{1}{m_{\mbox{\tiny Proca}}}=\frac{1}{e\sqrt{\sum_i u_i^2}}\,.
\end{equation}
Length scales associated with condensate's degrees of freedom are obtained in a more complicated way. 
Indeed they are given by the eigenvalue spectrum of a system of $6$ coupled (static) Klein-Gordon equations, whose 
(squared) mass matrix $\mathcal{M}^2$ is derived from \Eqref{order2}. It may be instructive to have this 
mass matrix explicitly.  First of all, let us remark that fluctuations can be separated in two groups, the `density amplitude' 
$\vec f =(f_1,f_2,f_3)^T$, and the `normalized phase amplitudes' $\vec \pi = (\pi_1,\pi_2,\pi_3)^T$. This mass matrix is a real 
symmetric matrix, which is not diagonal and whose eigenvalues are the (squared) masses of the normal modes. 
The eigenspectrum of $\mathcal{M}^2$, defines the (squared) masses of the physical modes. The inverses of 
each of the masses gives the characteristic length scales of the theory. For example in a single component 
theory the inverse mass of the fluctuations of the modulus of the order parameter $|\psi |$ is the coherence 
length (up to a factor of $\sqrt{2}$). In a two-component theory the fluctuations in 
the  phase difference (the Leggett's mode) are  characterized by a mass, the inverse of which sets the length 
scale at which a perturbed phase difference recovers its ground state values. In two-component models the
 density modes are mixed: \ie the characteristic length scales of the density fields are associated with the linear 
combinations of the fields \cite{GL2mass,GL2mass2,Silaev}. Physically this means that disturbing one density 
field necessarily perturbs the other. It also implies that, say in a vortex, the long-range 
asymptotic behavior of both density fields is governed by the same exponent, corresponding to a mixed
mode with the lowest mass.

We see that in the three-component case a new situation can arise where different collective modes are possible which are associated with mixed density and phase modes
In the basis $(\vec f,\vec \pi)$, the (squared) mass matrix can be written in terms of $4$ sub-matrices 
\begin{equation}
   \gamma^{(1)}\mathcal{M}^2\gamma^{(1)}=(\vec f,\vec \pi)
   \MatrixTwo{M_{ff}&M_{f\pi}\\M_{\pi f}&M_{\pi\pi}}\Vector{\vec f\\\vec \pi}. 
\end{equation}
Where $M_{ff}$ and $M_{\pi\pi}$ are the self-coupling of density and phase fluctuations, 
while $M_{f\pi}$ and $M_{\pi f}$ blocks control the mixing of density modes and phase modes.
\begin{widetext}
\begin{align}
   M_{ff}&=\MatrixThree{
\alpha_1+3\beta_1u_1^2	& \bar\eta_{12}			&\bar\eta_{13} 		\\
\bar\eta_{12}				&\alpha_2+3\beta_2u_2^2	&\bar\eta_{23} 	\\
\bar\eta_{13}				&\bar\eta_{23}			&\alpha_3+3\beta_3u_3^2 
} \,,	\ \ \ \ \ \ \ 
 M_{\pi\pi}=\MatrixThree{
-\frac{u_2\bar\eta_{12}+u_3\bar\eta_{13}}{u_1}	&	\bar\eta_{12}	&	\bar\eta_{13}	\\
\bar\eta_{12}	&	-\frac{u_1\bar\eta_{12}+u_3\bar\eta_{23}}{u_2}	&	\bar\eta_{23}\\
\bar\eta_{13}&	\bar\eta_{23}	&	-\frac{u_1\bar\eta_{13}+u_2\bar\eta_{23}}{u_3}	
}	\,, \nonumber\\
 M_{f\pi}& =  M_{\pi f}^T=\MatrixThree{
-\frac{\hat\eta_{12}u_2+\hat\eta_{13}u_3}{u_1}	&	\hat\eta_{12}	&	\hat\eta_{13}	\\
-\hat\eta_{12}	&	\frac{\hat\eta_{12}u_1-\hat\eta_{23}u_3}{u_2}	&	\hat\eta_{23}	\\
-\hat\eta_{13}	&	-\hat\eta_{23}	&	\frac{\hat\eta_{13}u_1+\hat\eta_{23}u_2}{u_3}
}\,,
\end{align}
\end{widetext}
where for having more compact expression we introduce new notations  $\bar\eta_{ij}=\frac{\eta_{ij}}{2}\cos\gsphasediff$, 
and $\hat\eta_{ij}=\frac{\eta_{ij}}{2}\sin\gsphasediff$.
Finally in order to derive the length scales associated with the condensate fluctuations, one has to diagonalize the 
matrix $\mathcal{M}^2$. Its eigenspectrum is the set of $6$ squared masses $\mc M_i^2$, whose corresponding 
lengths $\ell_i=1/\sqrt{2}\mc M_i$ are the physical length scales of a three band superconductor. 
In appendix \ref{Units} we also show how these length scales are expressed in different units.
There is a spontaneously broken $U(1)$ symmetry associated with the simultaneous equal changes of all phases. 
The mass of this mode is zero and the  eigenvector associated with this $U(1)$ zero mode  can easily be decoupled.
Thus one can  reduce the size of the system. However we prefer not to decouple this mode from the mass spectrum, 
since it provides a measure of the error of the numerical resolution of masses of other modes.
 The corresponding degree of freedom is described by the first term in \Eqref{LondonEnergy},
it is a  $U(1)$ Goldstone boson which, due to its coupling to the gauge field $\bf A$ yields a massive vector 
field with the mass $m_{\rm Proca}$ via the Anderson-Higgs mechanism.

Unfortunately the eigenbasis of $\mathcal{M}^2$ cannot be known analytically, in the 
general case. We calculate it numerically below. 
\subsection{Numerical results}

\begin{figure*}[!htb]
 \hbox to \linewidth{ \hss
\includegraphics[width=\linewidth]{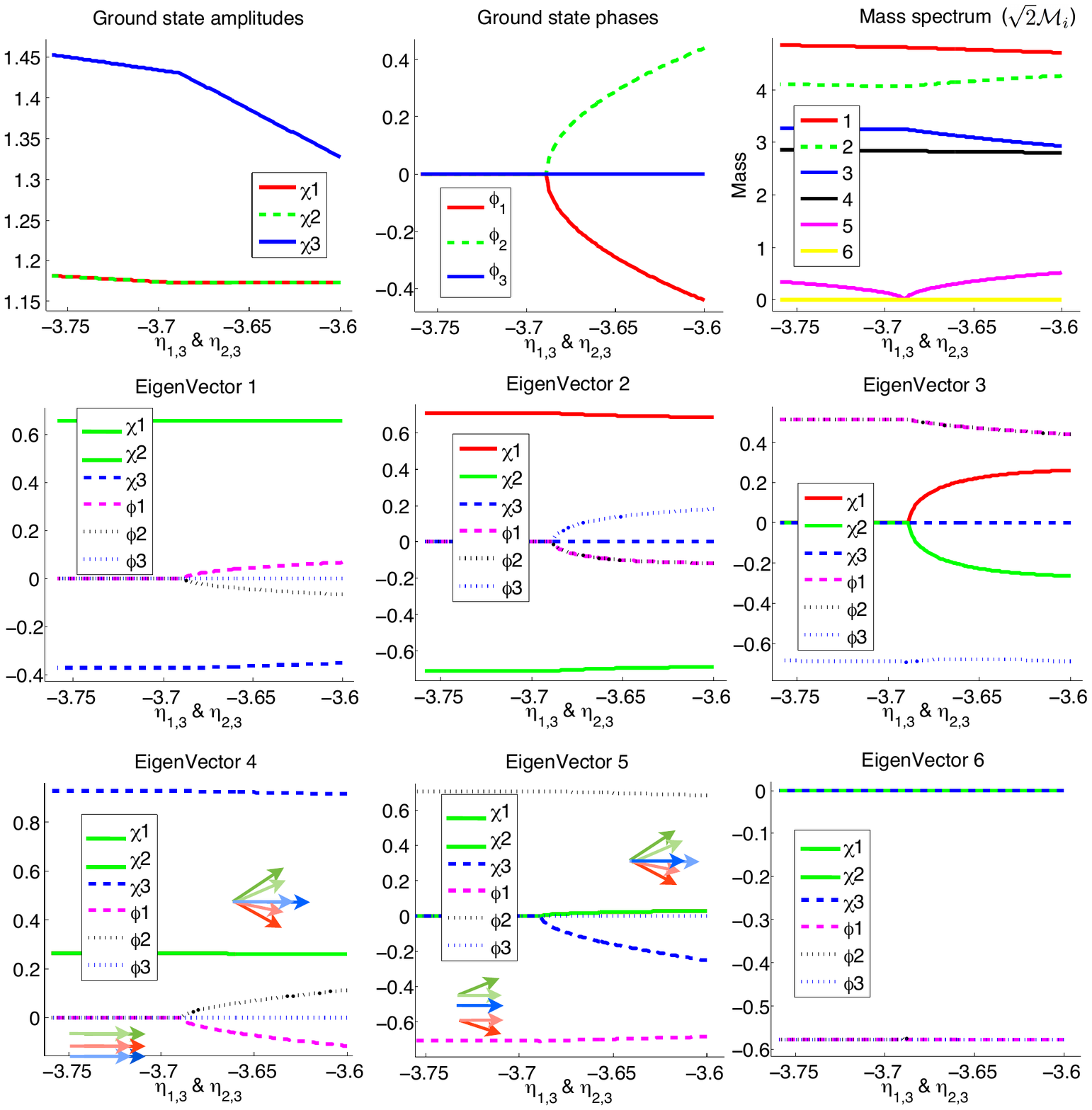}
 \hss}
\caption{
(Color online) Ground state field values, eigenspectrum and eigenvectors of the mass matrix. The x-axis gives the two parameters 
$\eta_{13}=\eta_{23}$ while the other parameters are 
$\alpha_1=-3,\;\beta_1=3, \;\alpha_2=-3,\;\beta_2=3, \;\alpha_3=2,\;\beta_3=0.5,\;\eta_{12}=2.25$. 
The eigenvectors are sorted according to corresponding eigenvalue, starting with the largest. The smallest eigenvalue is the zero mode associated with the spontaneously broken U(1) symmetry (no coupling to gauge field $\bf{A}$ is implied here, which otherwise renders this mode massive via  Andreson-Higgs mechanism).
At $\eta_{13}=\eta_{23}\approx -3.69$ there is a transition from a $U(1)$ to a $U(1)\times Z_2$ state.
The eigenvalue (5) becomes zero at the transition point, so there appears a divergent length scale at this point 
which correspond to the eigenvector (5)  \ie the phase difference mode becomes a scaleless 
collective excitation. Observe that in the $U(1)$ region, the eigenvectors exhibit no mixing between densities and phases, whilst 
in the $U(1)\times Z_2$ region there is in fact not a single eigenvector that does not exhibit mixing of phases and densities. 
Then, perturbation of densities are generically associated with perturbations of the phase differences in this regime. 
The arrows on the panels in the bottom row illustrate the variation of
the fields associated with the corresponding collective modes. Length of the arrows corresponds to modulus fluctuation and the direction of the
arrow corresponds to phase fluctuation. Modulus field of component 1 in red, component 2 in green and component 3 in blue. 
}
\label{eigen}\label{bifurc}
\end{figure*}

\Figref{eigen} shows the ground state, eigenspectrum and eigenvectors of the (squared) mass matrix in a frustrated three-band
superconductor as a function of the Josephson couplings. The coupling $\eta_{12}$ is fixed, while the 
horizontal axis gives the coupling coefficients $\eta_{13}$ and $\eta_{23}$. Each eigenvector is a 
linear combination of the degrees of freedom that comprises  a physical mode,   which variation length scale is given
by the square root of the inverse of the corresponding eigenvalue in the eigenspectrum. The system crosses 
over from $U(1)$ to $ U(1)\times Z_2$ TRSB state  at $\eta_{13}=\eta_{23}\approx -3.69$. 
Interestingly, in the $U(1)$ regime, the density modes are mixed. However, 
as can be seen from the eigenvectors, there is no mixing between density modes and the phase modes. 
Thus, perturbations of the densities and of the phases recover independently from each other. 
The fluctuations of the phase modes are the three-component generalization of the standard Leggett's modes. 
In the $U(1)\times Z_2$ regime the situation is opposite, and all eigenvectors are mixed in density and phase. 
This indicates that any perturbation of the densities creates a perturbation to the phases, and vice versa. 

There is a point where a Leggett's mode becomes massless, as was also pointed out recently in the phase-only model in \cite{hu2}. 
This occurs at the transition from $U(1)$ to $U(1)\times Z_2$ (note however, that the transition between these states can be first order as discussed in \cite{stanev}). In \Figref{eigen} panel (c) the eigenvalue 5 does indeed go to 
zero, indicating that the mass vanishes. The corresponding eigenvector can be seen in panel (h). In the $U(1)$ 
regime it corresponds to perturbation of the phases 1 and 2. The physical implication is  that the recovery 
of a perturbation at this point is governed not by an exponential, but by a power law. It is only a point in 
the parameter space where this mass is zero. However there is a finite area in the parameter space around that point where 
although the mode is massive, the length scale associated with this it is anomalously large as a consequence of the frustration between 
Josephson couplings.

\section{Vortex matter in  three-band type-1.5 superconductors}

\subsection{Topological defects in three-band Ginzburg-Landau model}

Lets us start with outlining the basic properties of the vortex excitations. In case of a $[U(1)]^3$  
Ginzburg-Landau model (\ie when $\eta_{ij}=0$)  there are three ``elementary" vortex excitations 
associated with $2\pi$ winding in only one of the  phases : $\oint_\sigma \nabla \varphi_i = 2\pi$, where $\sigma$ 
is a closed path around a vortex core. Such a vortex carries a fraction of flux quantum as can be seen from 
the following argument \cite{frac,smiseth}: the supercurrent  in case when there is a phase winding in only one phase is
\bea
{\mbf J}_i= \frac{ie}{2}[\psi_i^* \nabla \psi_i-\psi_i\nabla \psi_i^*] -
e^2\sum_k|\psi_k|^2{\mbf A}
\eea
Expressing $\mbf A$ via gradients and choosing the contour $\sigma$ far from the vortex core gives the following 
equation for the magnetic flux
\bea
\Phi_i&=&\oint_\sigma {\mbf A} d{\mbf
l}=\frac{u_i^2}{\sum_{k=1,2,3}u_k^2}\frac{1}{e}\oint_\sigma \nabla
\phi_i \nonumber \\
&=&\frac{u_i^2}{\sum_{k=1,2,3}u_k^2}\Phi_0
\label{flux}
\eea
where $\Phi_0$ is a flux quantum. Such a fractional vortex in the $[U(1)]^3$ case has logarithmically divergent
energy. Thus in external field a bulk three-component superconductor should form ``composite" integer flux
vortices which have phase winding in all components:  
$\oint_\sigma \nabla \varphi_1 = 2\pi,\oint_\sigma \nabla \varphi_2 = 2\pi,\oint_\sigma \nabla \varphi_3 = 2\pi$.
When  Josephson coupling is non-zero, then energy of a fractional vortex diverges linearly \cite{frac} and thus 
a single integer flux vortex in a bulk superconductor can be viewed as a strongly bound state of three co-centered
fractional flux vortices. 
Note that such a bound state will in general have three different sizes of vortex
cores. The characteristic length scales of the density recovery in the vortex cores are determined by the inverse 
masses of normal modes calculated above. Note also that the role of Josephson interaction on vortices is different  
in the presence of domain walls in three-band $U(1)\times Z_2$ superconductors.  Immediately at the domain wall the
Josephson terms have energetically unfavourable values of the phase differences. Thus, if a composite vortex is placed 
on such a domain wall, the Josephson interaction can force  a splitting of this vortex into fractional flux vortices, 
because the splitting will
allow to attain a more favorable configuration of the phase differences \cite{skyrmions}.

\subsection{Qualitative argument on the vortex states in frustrated three-band superconductors}

The ground state of a phase frustrated superconductor is in many cases non-trivial, with phase differences being 
compromises between the various interaction terms. Inserting vortices in such a system can shift the balance between
different competing couplings, since
vortices can in general have different effects on the different bands. In particular, since the core sizes of vortices 
are not generally the same in all bands, vortex matter will typically deplete some components more than others and 
thus can alter the preferred  values of the phase difference. So the minimal potential energy inside a vortex lattice or 
cluster may correspond to a different set of phase differences than in the vortexless ground state.
In this subsection we give a qualitative description of it, using an ansatz-based
argument. In the following section we study this question numerically without involving an ansatz.

The qualitative argument is as follows. Consider the phase-dependent potential terms in the free energy \Eqref{freeEnergy} which are of the form
\bea
\eta_{ij} u_iu_j f_i(\mbf r)  f_j(\mbf r)\cos(\varphi_{ij}(\mbf r))\,,
\eea
where $u_i$ are ground state amplitudes and each $f_i(\mbf r)$ represent an ansatz which models how superfluid densities 
are modulated due to vortices. Consider now a  system where N vortices are uniformly distributed in a domain $\Omega$. 
The phase dependent part of the free energy is 

\bea
U_\varphi=\left[\sum_{i>j}\eta_{ij} u_iu_j \right]
  \int_{\Omega} d\mbf r  f_i(\mbf r)  f_j(\mbf r)\cos(\varphi_{ij}(\mbf r)).\label{fullrenorm}
\eea
If $\varphi_{ij}$ is varying slowly in comparison with the inter vortex distance, then it can be considered
constant in a uniform distribution of vortices (as a first approximation). In that case \Eqref{fullrenorm} can be 
approximated by
\begin{equation}
U_\varphi\simeq\sum_{ij}\tilde\eta_{ij} u_iu_j\cos(\varphi_{ij})~\text{where}~\tilde{\eta}_{ij}=\eta_{ij}\int_{\Omega} d\mbf r  f_i(\mbf r)  f_j(\mbf r)
\label{renorm}
\end{equation}
If on the other hand $\varphi_{ij}$ varies rapidly, then it is not possible to define $\tilde{\eta}_{ij}$ without a spatial 
dependence. Then $\varphi_{ij}$ will depend on $\tilde{\eta}_{ij}(\mbf r)$ which is related to the local modulation functions 
$f_if_j$ and vary with a length scale given by the mass matrix \Eqref{order2}.

Thus, $\tilde{\eta}$ is the effective inter-band interaction coupling resulting from density modulation. 
Since in general, $f_i\not=f_j$ (unless the two bands $i,j$ are identical), one must take into account the 
modulation functions $f_i$ when calculating the phase differences. In particular, if the core size in 
component $i$ is larger than in component $j$, then  $\int d\mbf r f_if_k <\int d\mbf r f_jf_k$ 
and therefore the phase differences $\varphi_{ij}$ minimizing \Eqref{renorm} 
depend on $f_i$, and consequently on the density of vortices. 
Roughly speaking, introducing vortices in the system is equivalent to relative effective decrease of some of the Josephson coupling
constants.

Because the problem is nonlinear, the modulation functions $f_i$ generally depend on $\varphi_{ij}$ 
since the vortex core shape depends on the inter band interactions. As a result, an exact solution to this problem can only be found by numerical methods.  Below, we address 
this problem  by finding numerically  vortex clusters solutions. 
Some qualitative statements can nonetheless be made about these systems:

\begin{itemize}
\item If band $i$ is associated with larger vortex cores than band $j$, then with increasing 
density of vortices, the effective Josephson coupling $\tilde{\eta}_{ik}$ is depleted faster than $\tilde{\eta}_{jk}$.

\item The average intercomponent phase difference in a vortex cluster depends on the parameters $\tilde{\eta}_{ij}$.  
So the intercomponent phase differences inside a vortex cluster can be different from the vortexless ground state. 
Superconductors with $U(1)\times Z_2$ symmetry and disparity of core  sizes will therefore generally exhibit  perturbation of 
the phase differences due to vortices.

\item The symmetry of the system depends on the inter-band interactions,  so
 vortex matter can induce a phase transition between $U(1)$ and $U(1)\times Z_2$ states or vice versa 
\end{itemize}
This physics  depends on the spatial distribution 
of vortices in the system. 

If vortices are uniformly distributed in the sample, 
as is generally the case in clean type-2 superconductors, then the effective inter-band 
interaction strengths $\tilde{\eta}_{ij}$ are depleted in the same way everywhere in the sample. 
A change in symmetry $U(1)\to U(1)\times Z_2$ would then occur in the 
whole system at a certain value of applied external field. 

It also opens a possibility of a type-1.5 regime qualitatively different  in three-band systems than 
in their two band counter parts. Indeed, because of the non-monotonic interactions, the superconductor 
possesses macroscopic separation of Meissner domains and vortex clusters. In the three-band case, these 
phases can exhibit different broken symmetries. For example, Meissner domains with the $U(1)$ symmetry 
and vortex clusters having a different symmetry, $U(1)\times Z_2$. The $U(1)\times Z_2$ broken symmetry arising here because of 
the renormalization by vortices of the effective coupling constants $\tilde{\eta}_{ij}$. 
If there is a symmetry change $U(1)\to U(1)\times Z_2$ associated with vortex clusters in the system then there will
be two kinds of vortex clusters corresponding to with $Z_2$ states. They will coexist with 
the Meissner states voids which do not have the broken $Z_2$ symmetry.
Clearly, because of  this additional discrete symmetry, inter-cluster interaction should generally be affected by whether 
the clusters are in the same, or in different $Z_2$ states. When the magnetic field increases vortex clusters 
will merge and the entire system will be in the  state with broken $U(1)\to U(1)\times Z_2$ symmetry.

\subsection{Numerical results}
We used numerical computations to examine the questions which were raised about vortex matter in the previous sections.
The free energy functional \Eqref{freeEnergy} is minimized in presence of vortex matter.
In these simulations the variational problem was defined using a finite element formulation provided 
by the Freefem++ \cite{Freefem} library framework, using a Nonlinear Conjugate Gradient method. 
Reader interested in more technical details can refer to the appendix \ref{Numerics}. From 
this numerical data, several observations about vortex matter in three-band systems can be made.

\subsubsection{\texorpdfstring{Vortex clusters with broken  $Z_2$ symmetry}{Vortex clusters with broken  Z2 symmetry}}

We have simulated vortex clusters in a type-1.5 regime in the system given in \Figref{bifurc} for $\eta_{13}=\eta_{23}=-3.7$, 
\ie in the $U(1)$ region but close to the transition to broken time reversal symmetry. 
Thus, if the vortex core size in component 3 is larger than in the bands 1 and 2, then we should expect 
the breakdown of time reversal symmetry, for a sufficiently high density of vortices. 
\Figref{ch1} shows that this is indeed the case. In the ground state, all phases are equal ($\bar\varphi_1=\bar\varphi_2=\bar\varphi_3$), 
but once vortex clusters are present, these phases are no longer preferable and two other equivalent phase locking states develops. 
As the density of the third band is depleted, phase differences come to be increasingly dominated by the interband coupling 
between the two other bands. This coupling term not being minimal for $\varphi_1=\varphi_2$. 

\subsubsection{Long-range intervortex forces}
Vortex matter in this system is associated with
substantial variations of the intercomponent phase differences. 
As discussed above, in three-band system there is a phase difference mode that becomes 
less and less massive  as we approach the transition to a TRSB state. Thus in the vicinity 
of this point the mass of the corresponding mode can be very small  and then
characteristic lengths of its variation, very large. 
This provides  an additional mechanism that can lead to vortex interactions at very large distances. 
\begin{figure*}[!htb]
\hbox to \linewidth{ \hss
\includegraphics[width=\linewidth]{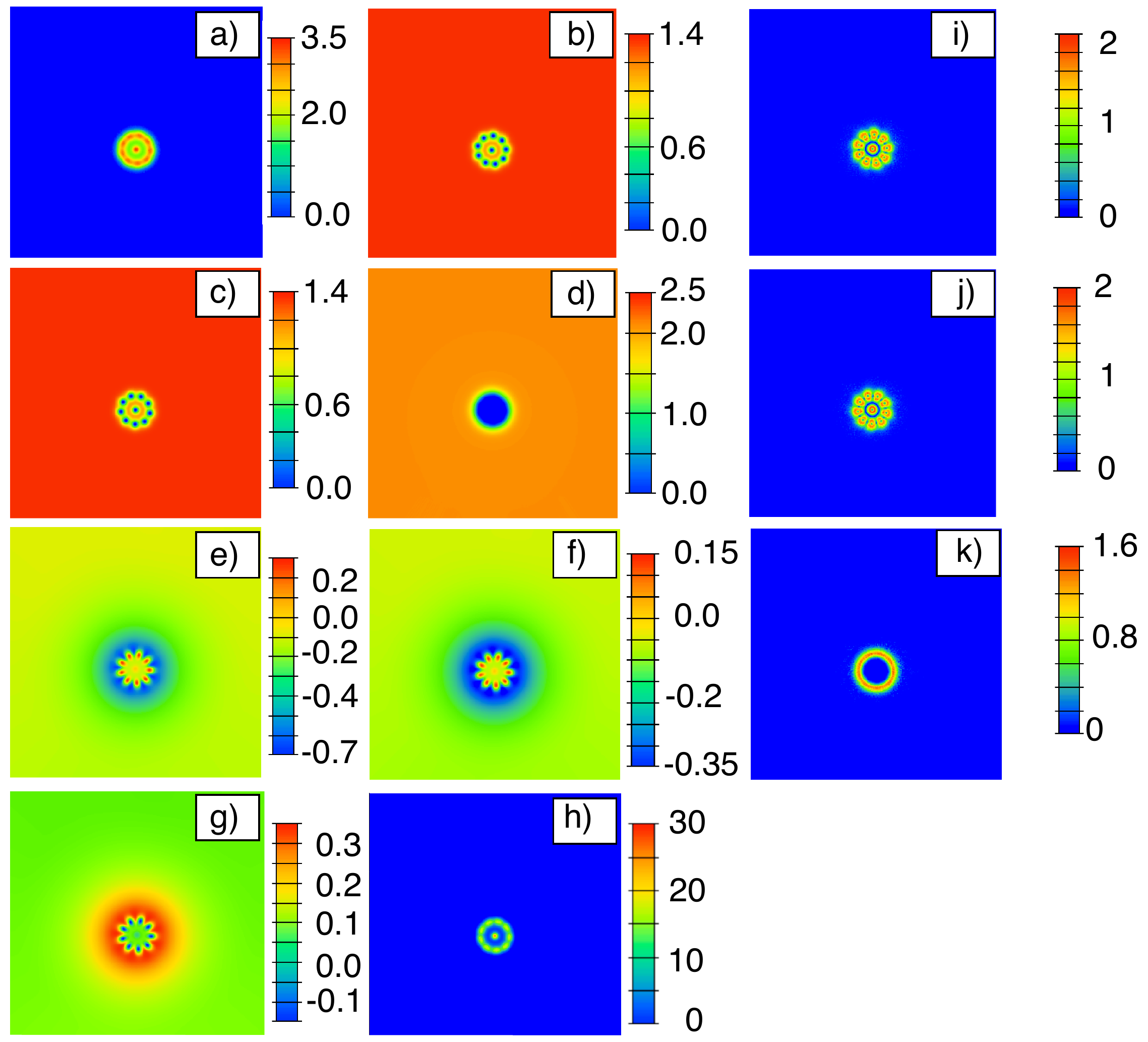}
 \hss}
\caption{
Vortex cluster exhibiting  an internal $Z_2$ state in a frustrated three-band superconductor. The panel 
displays the following quantities.
a) is the magnetic field, and the densities of the different condensates are b) $|\psi_1|^2$, c) $|\psi_2|^2$, d) $|\psi_3|^2$.
To monitor the relative phase differences, we use
e) $|\psi_1||\psi_2|\sin(\varphi_{12})$, f) $|\psi_1||\psi_3|\sin(\varphi_{13})$, g) $|\psi_2||\psi_3|\sin(\varphi_{23})$. Panel 
h) is the energy density and the supercurrents in each condensate are displayed on i) $J_1$, j) $J_2$ , k) $J_3$. 
The GL parameters used for this simulation are  
$\alpha_1=-3,\;\beta_1=3, \;\alpha_2=-3,\;\beta_2=3, \;\alpha_3=2,\;\beta_3=0.5,\;\eta_{12}=2.25,\;\eta_{13}=-3.7,\;\eta_{23}=-3.7$. 
Thus, they correspond to the $U(1)$ region in Fig. \ref{eigen}, but close to the transition point to $U(1)\times Z_2$ symmetry.  
In the ground state, all the phases are locked ($\bar\varphi_1=\bar\varphi_2=\bar\varphi_3$) 
as a consequence of the Josephson couplings $\eta_{12}=\eta_{13}=-3.7$ dominating the interaction. 
Inside the vortex cluster the third condensate is depleted, so the coupling terms 
$\eta_{i3}|\psi_i||\psi_3|\cos(\varphi_{i3}),\;\{i=1,2\}$ 
become much weaker while the term $|\psi_1||\psi_2|\eta_{12}\cos(\varphi_{12})$ becomes dominant. In sufficiently dense 
vortex matter, the ground  state is changed due to the dominating antilocking interaction between 
the components 1 and 2. This results in a $U(1)\times Z_2$ 
state inside the vortex cluster, as can be seen from the phase differences plots shown in panels e), f),g). (
Note that in the very center of the vortex cluster this quantity is small because of small values of the prefactors $|\psi_i||\psi_j|$.) 
A closer inspection of the density panels b) and c) reveals that vortex cores in both densities do 
not necessarily superimpose (it can also be seen from the supercurrents on panels i and j) and 
so they are fractional vortices.
This fractionalization occurs at the boundary of the cluster, while the vortex in the middle is a composite one-quantum vortex.
}
\label{ch1}
\end{figure*}
\Figref{ch2} displays the same system as in \Figref{ch1}, but with two vortex
clusters rather than one. A clearly visible perturbation of the phase differences 
extends from the clusters well outside the region with magnetic field
and far beyond the area with significant density suppression, providing a mechanism for 
long range inter cluster interaction.  

\subsubsection{Vortex fractionalization in clusters}

\begin{figure*}[!htb]
\hbox to \linewidth{ \hss
\includegraphics[width=\linewidth]{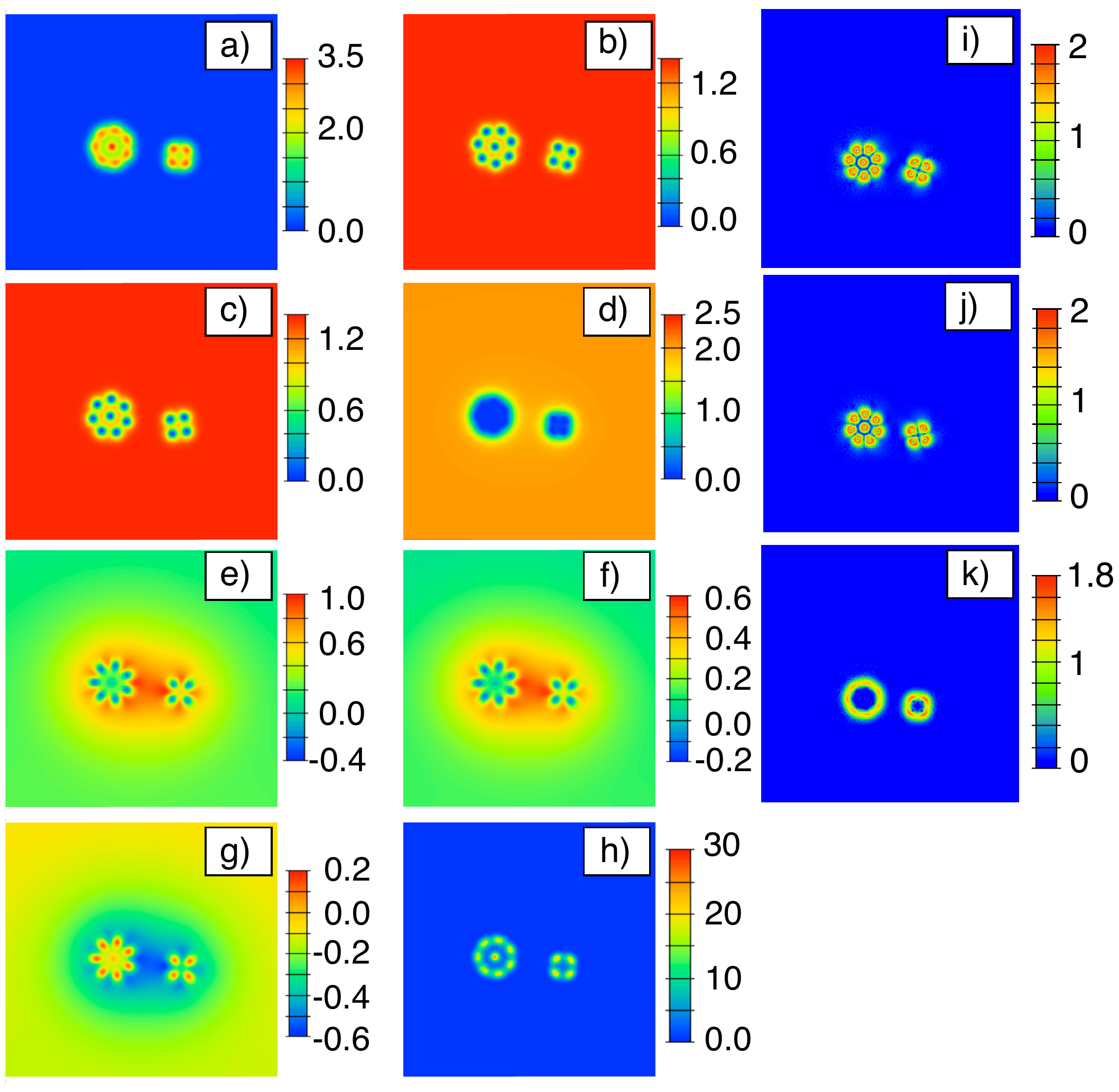}
 \hss}
\caption{
Interacting vortex clusters with internal $Z_2$ symmetry in a frustrated superconductor. The snapshot
represents a state where the energy is well minimized with respect to all variables except the relative positions of the 
weakly interacting well-separated clusters. 
The GL parameters and displayed quantities in the panels are the same as  in \Figref{ch1}.
The analysis of the eigenvalues in \Figref{eigen} shows that 
 there a mode  with a very small mass,  associated with the eigenvector $[0,0,0,1,-1,0]$.
It corresponds to the mode which associated with phase
difference fluctuations and it has the largest recovery length scale.
This is indeed visible in the plots e), f), and g). The phase difference $\varphi_{12}$ 
(e) recovers much more slowly than the magnetic field (a) and the condensate densities (b-d). 
Clusters clearly interact at a distance greatly exceeding the length scales of density modulation 
and the magnetic penetration depth, as this mode stretches out between them.
 }
\label{ch2}
\end{figure*}

\Figref{ch1} and \Figref{ch2} also exhibit flux fractionalization. 
As previously mentioned, the model \Eqref{freeEnergy} allows fractional vortex solutions, where 
only one of the phases $\varphi_i$ winds $2\pi$ around some point while the rest doesn't. 
The flux carried by a single fractional vortex is given by \Eqref{flux}.
Two forces hold fractional vortices together as a one-quantum composite vortex, in the three-component model. 
First is the interaction with the gauge field, which gives logarithmic interaction at long range \cite{frac,smiseth}. 
The second is the Josephson coupling, which is asymptotically linear. In non-frustrated superconductors the Josephson coupling gives attractive 
interaction between fractional vortices, but in frustrated systems this interaction can be repulsive, resulting in fractionalization of vortices 
\cite{skyrmions}. 

Consider the system in \Figref{ch1} and \Figref{ch2}. The ground state corresponds to 
$\bar\varphi_1=\bar\varphi_2=\bar\varphi_3$. Since there is an energy 
cost associated with gradients of the phase difference, these are expected to change slowly. Thus, far away from the cluster, the state is 
simply the ground state. Deep inside the cluster phase differences attain a broken $U(1)\times Z_2$ state, depending on the density of 
vortex matter. If the vortex density is very high, then $|\psi_3|$ is very small, and we expect inside the cluster 
$\varphi_{12}\to \pi$ (provided that the cluster is large). 
While $\varphi_{12}$ varies slowly, the density in $|\psi_3|$ recovers more rapidly at the boundary of the cluster. 
Thus, there may be an area where $|\varphi_{12}|<\pi/2$ while $|\psi_3|$ is small. 
Consequently the interaction between fractional vortices in the bands 1 and 2 due to Josephson coupling is repulsive in this area. 
Also when the magnitude $|\psi_3|$ is very small or zero, the Josephson inter-band coupling $\psi_{23}$ 
and $\psi_{13}$ which provides attractive interaction between  the fractional vortices is weaker or essentially disappears.
Thus, the interaction of the fractional vortices is governed by the coupling to the gauge field, which gives attractive interaction, 
and the remaining Josephson coupling, which in this case gives repulsive interaction. 
As a result, in that region the integer flux vortices split into fractional ones.
 
This effect is found in numerical simulations of vortex clusters. 
Looking carefully at \Figref{ch1} and \Figref{ch2} we can see that  the vortex
cores  in the bands 1 and 2 do not generally coincide. From panel g) we 
can read off that, at the boundary of the cluster, the phase difference between the components 1 and 2 is given by 
$|\psi_1||\psi_2|\sin(\varphi_{12})\approx -0.7\to |\varphi_{12}|\approx 0.5\to \cos(\varphi_{12})\approx \sqrt{3}/2>0$. 
Thus, in that region, the Josephson term associated with components 1 and 2 gives a positive energy contribution 
resulting in repulsive interaction between fractional vortices in components 1, 2 leading to fractionalization of vortices.
Indeed, fractionalization occurs for all vortices expect those in the center of the large 8 or 9 quanta clusters.
We observe in large systems, that fractionalization is important at the boundary of the clusters and 
becomes less pronounced for vortices located deep inside. The magnetic field is significantly smeared 
out, as a result of this fractionalization.

The fractionalization at the cluster's boundary has a similar origin as the 
physics which stabilizes topological solitons in the TRSB states in three-band superconductors
\cite{skyrmions}. The difference is however that the topological solitons discussed
in \cite{skyrmions} are stable bound states of $Z_2$ domain walls and fractional
vortices, while here there is not a $Z_2$ domain wall, but fractionalization comes
as a result of complicated behaviour of the fields at a cluster boundary which 
is an interface between $U(1)$ and $U(1)\times Z_2$ states.

\subsubsection{\texorpdfstring{$\pi$-walls}{Pi-walls}}

\begin{figure*}[!htb]
\hbox to \linewidth{ \hss
\includegraphics[width=\linewidth]{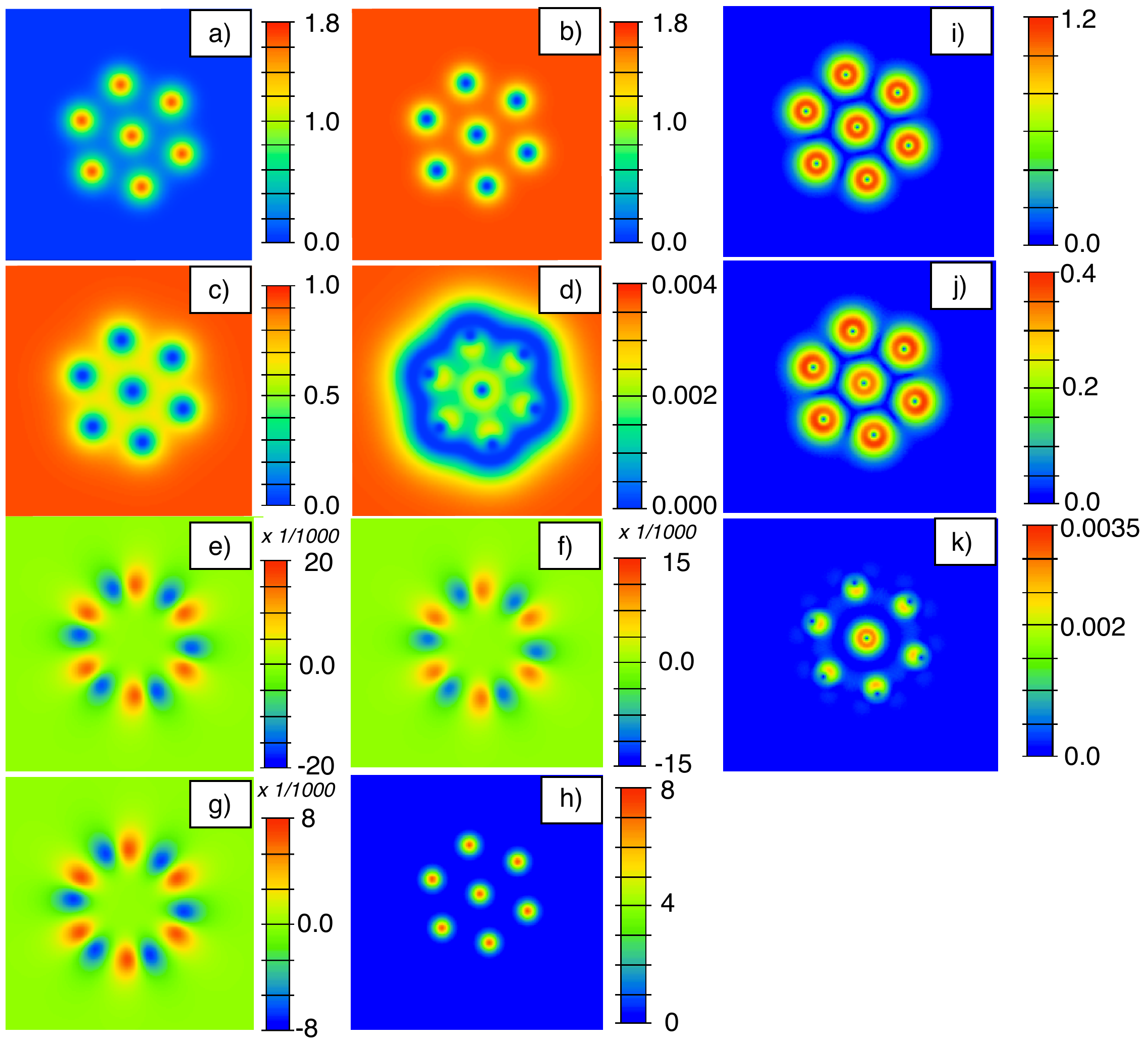}
 \hss}
\caption{
A vortex cluster surrounded by a $\pi$-wall. Here again displayed quantities and panel labels are the same as in \Figref{ch1}. The GL parameters are  
$\alpha_1=-1,\;\beta_1=1, \;\alpha_2=1,\;\beta_2=0.5, \;\alpha_3=3,\;\beta_3=0.5,\;\eta_{12}=-2,\;\eta_{13}=2.7,\;\eta_{23}=-4$.
In the ground state, the phases are locked ($\bar\varphi_1=\bar\varphi_2=\bar\varphi_3$), 
but frustration occurs as $\eta_{13}=2.7$ 
gives an antilocking interaction (i.e. the term $|\psi_1||\psi_3|\eta_{13}\cos(\varphi_{13})$ is minimal for 
$\varphi_{13}=\pi$). As vortices are introduced in the system, the superfluid densities are depleted. It is clear 
from visual inspection that the vortices in the second band are larger than those of the first. Thus, the effective coupling 
$\tilde{\eta}_{23}$ decreases faster than $\tilde{\eta}_{13}$ and so inside the vortex cluster the preferred phase 
becomes $\varphi_1=\varphi_2=\varphi_3+\pi$.  Since the third band has much smaller density than the other bands, the 
energetically cheapest way of coping with this is to create a domain wall-like object where $|\psi_3|$ becomes very small. 
It does not cost much energy to have a large phase gradient density, so that $\psi_3$ quickly picks up a $\pi$-shift in its phase.
As a result the density of  $|\psi_3|$ is suppressed not only in the vortex cores but also in a ring surrounding the vortex cluster as can be clearly seen in panel (d).
}
\label{piwall}
\end{figure*}

Another phenomena associated with frustrated superconductors are
objects which we term ``$\pi$-walls". In certain parameter regions, vortices and vortex clusters are surrounded by a 
domain-wall-like object with substantially suppressed density across which the phase of one of the 
condensates jumps by $\pi$.

\begin{figure*}[!htb]
 \hbox to \linewidth{ \hss
\includegraphics[width=\linewidth]{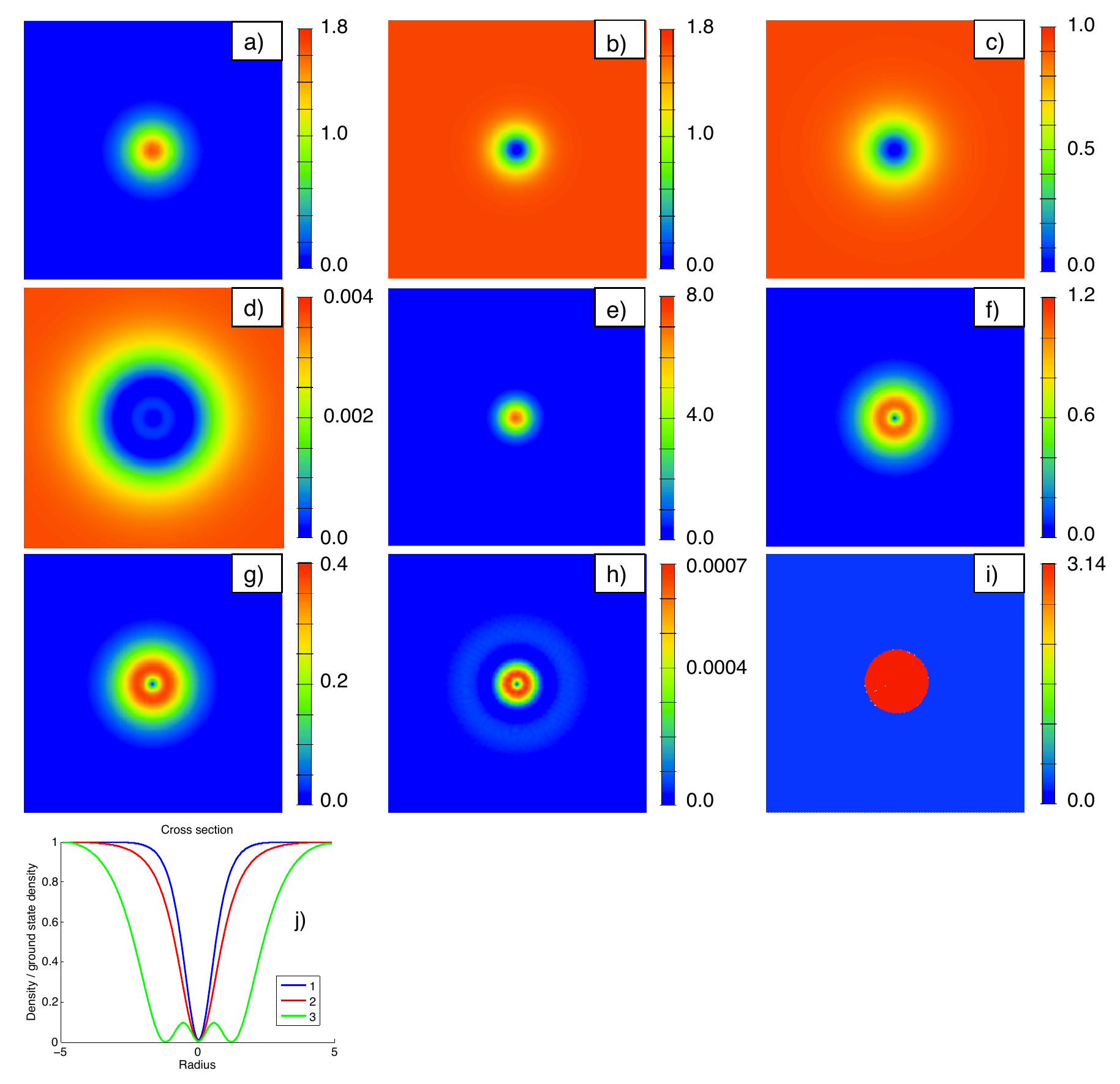}
 \hss}
\caption{ 
A single vortex in a system exhibiting $\pi$-wall solutions. 
The interband coupling coefficients are $\eta_{12}=-2,\; \eta_{13}=2.7,\; \eta_{23}=-4$. 
Displayed quantities are here the magnetic flux a) and densities of each condensate  b) $|\psi_1|^2$, c) $|\psi_2|^2$, 
d) $|\psi_3|^2$. Panel e) shows the total energy density while f), g) and h) are the supercurrents
( $J_1$, $J_2$ and $J_3$ respectively). The panel i) shows the phase difference $\varphi_{13}$. 
The parameters of the system are identical to the one shown on \Figref{piwall}. 
The $\pi$-wall can be seen from the double dip in density of the third band as can be seen in j), 
as well as from the phase difference plotted in i). Thus, $\psi_3$ is zero in the center, it recovers slightly and then drops close to zero 
again on a circular area at certain distance from the vortex center. At the second drop  the phase $\varphi_{3}$  picks up an extra 
phase $\pi$ as can be seen from the plot of $\varphi_{13}$on the panel i). 
}
\label{piwall1}
\end{figure*}

An example of such an object is displayed in \Figref{piwall}. The density in the third band is small in comparison 
to the other bands. The Josephson coupling $\eta_{12}=-2$ results in locked phases $\varphi_{12}=0$. The system 
is frustrated, since $\eta_{23}=-4$, preferring phase locking with respect to $\varphi_{23}$, and $\eta_{13}=2.7$ 
preferring phase antilocking with respect to $\varphi_{13}$. When there are no vortices in the system, the term 
$|\psi_2||\psi_3|\eta_{23}\cos \varphi_{23}$ dominates over $|\psi_1||\psi_3|\eta_{13}\cos \varphi_{13}$, and the 
ground state is $\bar\varphi_1=\bar\varphi_2=\bar\varphi_3$. 
However, when vortices are present in the system, this is not 
necessarily the case. The vortex cores in the second band are larger than those of the first, and consequently, 
the effective coupling strength $\tilde{\eta}_{23}$ is diminished at a higher pace than $\tilde{\eta}_{13}$. 
Thus, inside a vortex cluster, the potential energy is minimal when $\varphi_{13}=\varphi_{23}=\pi$. 
To comply with these requirements, the system forms a domain wall-like object where $|\psi_3|$ goes close to zero, 
and $\varphi$ picks up an extra phase of $\pi$. 
For this particular set of parameters, this in fact happens even for a single vortex, as can be seen in \Figref{piwall1}.

\section{Conclusions}

Recently there has been a growing interest to three-band superconductivity sparked by the discovery 
of the iron pnictide superconductors. The precise information about
the characteristic parameters for these materials is 
not know yet. Also the current experiments suggest that physics of vortex ordering patterns 
in currently available samples  is substantially affected by strong pinning 
\cite{viktor2,eskildsen1,esk3,hoffman,luan,beena}. We presented here a general study showing that in a three-band system there are many 
phenomena which are not present in  two-band models. As was previously observed
\cite{agterberg,nagaosa,stanev,tanaka,hu,hu2}, in the presence of more than two bands, a system 
can exhibit frustration between different competing interband Josephson terms. We considered
 possible physical consequences using  a three-band Ginzburg-Landau model.
To observe this physics in experiment in fact does not necessary require a three-band
superconductor but it would be sufficient to have a superconductor with phase anti-locking Josephson 
interaction (\ie $\eta>0$). Then as was observed in \cite{nagaosa,stanev} a phase-frustrated state can be induced 
in a Josephson coupled bi-layer made of this and singe-band superconductors.
In that case of the Josephson couplings is just real-space interlayer coupling. Thus it provides an opportunity to tune its value. 

We discussed that this can result in the appearance of modes with very long characteristic 
length scales even when the interband Josephson coupling is strong. Here we also discussed  that in the TRSB 
$U(1)\times Z_2$ state of the three-band Ginzburg-Landau model there are no ``phase-only" 
Leggett's modes, but instead the system has different mixed phase-density collective modes which involve both 
phase and density fluctuations. The physics of the coupled modes and associated different length scales 
substantially affects vortex matter in the system. The vortices can interact at distances much larger than the
length scale of  magnetic field localization or length scale at which most of the condensate density is 
recovered, because of the existence of slowly varying phase difference and low-mass mixed density modes.
This can give rise to non-monotonic intervortex interaction and type-1.5 regimes 
in systems  where it would not be expected. In particular, if a large-$\kappa$ parameter is estimated from 
the second critical field of the system, then this does not prohibit the existence of modes with length scales that
substantially exceed the penetration depth even at strong Josephson coupling.

Moreover the competing interactions can qualitatively affect the vortex structure as well. We showed the 
existence of vortex solution where density is suppressed not only in the core but also takes a second dip in some belt-like area around the vortex core or around the vortex cluster. Such features 
can in principle be detected in STM measurements.

Furthermore we showed that subjecting a three band system to an external field which induces vortices can shift the 
balance in competing interactions and result in change of the ground state symmetry. In type-2 systems where 
vortices are uniformly distributed, changes in the phase difference will also occur  quite  uniformly 
there could be   a phase transitions between $U(1)$ and $U(1)\times Z_2$ states resulting from an applied magnetic field. 
In the case of type-1.5 superconductivity, systems will feature not only macroscopic phase separation between 
vortex clusters and domains of Meissner state, but also  a macroscopic phase separation between the domains 
of $U(1)$ and $U(1)\times Z_2$ ground states. The transition from the semi-Meissner to vortex 
states in that case will then be associated with change of the symmetry from $U(1)$ to $U(1)\times Z_2$.

This work is supported by the  
Swedish Research Council, and by the Knut and Alice Wallenberg
Foundation through the Royal Swedish Academy of Sciences fellowship and by
NSF CAREER Award No. DMR-0955902. 


\clearpage
\appendix

\section{Unit system} \label{Units}

The Ginzburg-Landau free energy \Eqref{freeEnergy}, is written after suitable rescaling. 
Below we give details of these rescalings, in order to define 
the various quantities in the usual dimensionful theory. In the following, let us denotes the usual dimensionful 
quantities with accentuated fonts. Consider the following
\begin{align}
   \df F &=\frac{\hbar^2  c^2}{4\pi} F\,, &
   \df \psi_i &=\sqrt{\frac{\df m  c^2}{4\pi}}\psi_i\,, &
   \df{\mbf A} &=-\hbar  c \bf A\,, \nonumber\\
   \df \alpha_i &=\frac{\hbar^2}{\df m} \alpha_i\,, &
   \df \beta_i &=\frac{4\pi \hbar^2}{\df m^2  c^2}\beta_i\,, &
   \df \eta_{ij} &=\frac{\hbar^2}{\df m}\eta_{ij}\,,  
\label{rescaling}
\end{align}
where $c$ is the speed of light and $\hbar$ the reduced Planck constant, then converting the 
free energy \Eqref{freeEnergy} to 
\begin{align} 
 \df F= &\frac{1}{8\pi}(\nabla \times \df{\mbf A})^2+ 
   \sum_{i =1,2,3}\frac{\hbar^2}{2\df m}|\df{\mbf D}\df\psi_i|^2 \nonumber \\
+&\sum_{i =1,2,3}\df\alpha_i|\df\psi_i|^2+\frac{1}{2}\df\beta_i|\df\psi_i|^4 \nonumber\\
+&\sum_{i =1,2,3}\sum_{j>i}\df\eta_{ij}|\df\psi_i||\df\psi_j|\cos(\varphi_{ij}) \,.
\label{freeEnergyDimensionful}
\end{align}
Here $\df{\mbf D}=\nabla-i\frac{e}{\hbar c}\df{\mbf A}$ and $\df m$ is the mass of the cooper pairs

This rescaling is also applied to the perturbative expansion \Eqref{Expansion} of the problem, so that 
the Klein-Gordon system becomes
\begin{equation}
\df E_{\mbox{\tiny Klein-Gordon}} \equiv  \frac{\hbar^2}{2\df m}(\df\gamma^{(1)\,\prime})^2 +
   \df\gamma^{(1)}\mathcal{M}^2\df\gamma^{(1)}\,
\label{KGDimensionful}
\end{equation}
and then, (dimensionful) length scales of the massive modes of the condensate are 
\begin{equation}
 \df \xi_i = \sqrt{\frac{2\hbar^2}{\df m}}\frac{1}{\mc M_i}\,.
\label{coherenceDimensionful}
\end{equation}
In the $U(1)\times Z_2$ regime, since all the mode are mixed, the length 
scales $\xi_i$ then are related to inverse masses of the modes.

London penetration depth is defined through the Proca equation of the gauge field, 
which reads now in the dimensionful system
\begin{equation}
\df E_{\mbox{\tiny Proca}} \equiv  \frac{1}{8\pi}(\frac{\df a^\prime}{r})^2+\frac{e^2\sum_i\df u_i^2}{2 \df m c^2}  \df a \,.
\label{ProcaDimensiomful}
\end{equation}
London penetration depth, which gives the exponential decrement of the magnetic field in the superconductor 
the reads
\begin{equation}
 \df \lambda^2 = \frac{\df m c^2}{4\pi e^2\sum \df u_i^2}\,.
\label{penetrationDimensionful}
\end{equation}

\section{Numerical Methods -- Finite element energy minimization} \label{Numerics}

We provide here a detailed description of the numerical methods which are used to construct vortex 
solutions in three component Ginzburg-Landau models. They are constructed by minimizing the 
free energy \Eqref{freeEnergy}, from an appropriate initial guess carrying several flux quanta.
We consider the two-dimensional problem $\mc F=\int_\Omega F \mathrm{d}x^2$ defined on the 
bounded domain $\Omega\subset\mathbbm{R}^2$, supplemented by `open' boundary conditions 
on $\partial\Omega$. This `open constraint' is a particular Neumann boundary condition, such that 
the normal derivative of the fields on the boundary are zero. These boundary conditions in fact are a 
very weak constraint. For this problem one could also apply Robin boundary conditions on $\partial\Omega$, so 
that the fields satisfy linear asymptotic behavior (exponential localization) \Eqref{Proca}. However, we choose to apply 
the `open' boundary conditions which are less constraining for the problem in question. `Open' boundary conditions also imply 
that vortices can easily escape from the numerical grid, since it would further minimize the energy. To 
prevent this, the numerical grid is chosen to be large enough so that the attractive interaction with the boundaries 
is negligible. The size of the domain is then much larger than the typical interaction length scales. Thus in this method 
one has to use large numerical grids, which is computationally demanding. At the same time the advantage 
is that it is guaranteed that obtained solutions are not boundary pressure artifacts. 

The variational problem is defined for numerical computation using a finite element formulation provided by the 
Freefem++ library \cite{Freefem}. Discretization within finite element formulation is done via a (homogeneous) 
triangulation over $\Omega$, based on Delaunay-Voronoi algorithm. Functions are decomposed on a continuous 
piecewise quadratic basis on each triangle. The accuracy of such method is controlled through the number of triangles, 
(we  typically used   $3\sim6\times10^4$), the order of expansion of the basis on each triangle (P2 elements being 2nd 
order polynomial basis on each triangle), and also the order of the quadrature formula for the integral on the triangles. 

Once the problem is mathematically well defined, a numerical optimization algorithm is used to solve the variational nonlinear 
problem (\ie to find the minima of $\mathcal{F}$). We used here a  Nonlinear Conjugate Gradient method. The algorithm is 
iterated until relative variation of the norm of the gradient of the functional  $\mathcal{F}$ with respect to all degrees of 
freedom is less than $10^{-6}$. 

\subsection*{Initial guess}
The initial field configuration carrying $N$ flux quanta is prepared by using an ansatz which imposes phase windings 
around spatially separated $N$ vortex cores in each condensates : 
\begin{align}
\psi_1&= |\psi_1|\mathrm{e}^{ i\Theta} \, ,
 \psi_2= |\psi_2|\mathrm{e}^{ i\Theta+i\Delta_{12}} \, ,
 \psi_3= |\psi_3|\mathrm{e}^{ i\Theta+i\Delta_{13}} \, ,~~  \nonumber \\
 |\psi_j| &= u_j\prod_{i=1}^{N_v} 
 \sqrt{\frac{1}{2} \left( 1+\tanh\left(\frac{4}{\xi_j}({\cal R}_i(x,y)-\xi_j) \right)\right)}\, ,~~  \nonumber \\
 \mbf A&=
 \frac{1}{e{\cal R}}\left(\sin\Theta,-\cos\Theta \right)\,,
 \label{InitialGuess1}
\end{align}
where $j=1,2,3\,$ and $u_j\,$ is the ground state value of each superfluid density. The parameter $\xi_j$ gives 
the core size while $\Theta\,$ and $\cal{R}\,$ are 
\begin{align}
\Theta(x,y)&=\sum_{i=1}^{N_v}\Theta_i(x,y) \,,  \nonumber\\
 \Theta_i(x,y)&=\tan^{-1}\left(\frac{y-y_i}{x-x_i}\right)\,, \nonumber\\
 {\cal R}(x,y)&=\sum_{i=1}^{N_v}{\cal R}_i(x,y)\,, \nonumber\\
 {\cal R}_i(x,y)&=\sqrt{(x-x_i)^2+(y-y_i)^2}\,. 
\label{InitialGuess2}
\end{align}
The initial position of a  vortex is given by $(x_i,y_i)$. The  functions $\Delta_{ab}\equiv\varphi_b-\varphi_a$ can be 
used to initiate a domain wall, when the ground state exhibits $U(1)\times Z_2$ symmetry.

\bibliographystyle{apsrev}


\end{document}